\documentclass{WileyMSP-template}

\begin{document}

\pagestyle{fancy}
\fancyhf{}
\fancyfoot[C]{\thepage}


\title{Controlled ion--ion interactions and cavity-enhanced emission of a coherent dinuclear Eu\textsuperscript{3+} complex}

\maketitle

\author{Evgenij Vasilenko\equalcontrib, Vishnu Unni Chorakkunnath\equalcontrib, Barbora Brachnakova\equalcontrib, Nicholas Lester Jobbitt, Senthil Kumar Kuppusamy, David Hunger*, and Mario Ruben*}


\dedication{$^{\dagger}$\ These authors contributed equally to this work.}

\begin{affiliations}

Evgenij Vasilenko, Barbora Brachnakova, Senthil Kumar Kuppusamy, David Hunger, Mario Ruben\\
Karlsruhe Institute of Technology, Institute for Quantum Materials and Technologies (IQMT), Eggenstein-Leopoldshafen, 76344, Germany\\
E-mail: \texttt{mario.ruben@kit.edu; david.hunger@kit.edu}\linebreak

Evgenij Vasilenko, Vishnu Unni Chorakkunnath, Nicholas Lester Jobbitt, David Hunger\\
Karlsruhe Institute of Technology, Physics Institute (PHI), Wolfgang-Gaede-Straße 1, Karlsruhe, 76131, Germany\linebreak

Mario Ruben\\
Karlsruhe Institute of Technology, Institute of Nanotechnology (INT), Eggenstein-Leopoldshafen, 76344, Germany\linebreak

Mario Ruben\\
Centre Europ\'{e}en de Sciences Quantiques (CESQ), Institut de Science et d’Ing\'{e}nierie Supramol\'{e}culaires (ISIS), Strasbourg, 67083, France\\

\end{affiliations}


\keywords{Molecular spin qubits, ensemble spectroscopy, rare-earth ions, optical coherence, dipole--dipole interactions, cavity spectroscopy}

\begin{abstract}

Molecular rare-earth-ion complexes offer unique opportunities for quantum technologies by combining the intrinsic coherence properties of rare-earth ions with chemically tunable molecular environments. A crucial capability is the realization of multi-qubit architectures with defined qubit couplings to enable two-qubit quantum gates. Here, we investigate the optical coherence properties and excitation-induced interactions of two \ce{Eu^{3+}}-based molecular complexes, comparing a mononuclear reference system with a dinuclear analogue in which two \ce{Eu^{3+}} ions are positioned at a well-defined intramolecular distance of about \SI{7}{\angstrom}. Using cryogenic ensemble spectroscopy, including spectral hole burning, free-induction decay, and photon echo measurements at temperatures down to \SI{100}{\milli\kelvin}, we demonstrate long optical coherence times $T_{2,\text{o}}$ of up to \SI{9}{\micro\second}. As a key step toward scalable multi-qubit architectures, a control-target sequence was implemented to probe conditional ion-ion interactions, revealing a stronger interaction-induced dephasing in the dinuclear complex. Finally, we show the integration of the dinuclear complex into a fiber-based optical microcavity, and observe an 380-fold emission enhancement of the $\mathrm{}^5\mathrm{D}_0\rightarrow\mathrm{}^7\mathrm{F}_0$ transition. Together, these results position molecular rare-earth complexes as versatile and chemically tunable building blocks for scalable quantum technologies.

\end{abstract}


\section{Introduction}
\label{Introduction}

Optically addressable solid-state spins have been established as a successful approach to realize quantum networks, implement quantum sensors, and demonstrate quantum processing nodes \cite{awschalom_quantum_2018,atature_material_2018,katsumi_recent_2025}. A remaining challenge is the development of scalable architectures of multiple qubits with defined couplings at high density.
Rare-earth ions (REI) doped in solid-state host materials have emerged as a promising material platform in this context, combining long-lived electronic and nuclear spin states with narrow optical transitions originating from the well-shielded $4f$ shell at cryogenic temperatures \cite{goldner_chapter_2015}. Optical detection and qubit control of individual \ce{Er^{3+}} \cite{dibos_atomic_2018,Ulanowski22_SciAdv,Deshmukh:23,gupta_dual_2023} and \ce{Yb^{3+}} ions \cite{kindem_control_2020} in crystalline hosts have been demonstrated using nanophotonic structures, enabling single-ion spectroscopy and coherent spin control of multiple qubits on the nanoscale \cite{chen_parallel_2020,ruskuc_multiplexed_2025}. In terms of spin coherence time $T_{2,\text{s}}$, \ce{Eu^{3+}}-doped materials are among the best, and ensemble experiments with \ce{Eu^{3+}}:\ce{Y2SiO5} have achieved spin coherence times extending up to \SI{18}{\hour} \cite{wang_nuclear_2025}, establishing \ce{Eu^{3+}} as an outstanding candidate for long-lived quantum memories. Furthermore, electric dipole-dipole interactions enable two-qubit quantum gates \cite{ohlsson_quantum_2002,longdell_demonstration_2004,kinos_designing_2021}, while another co-doped REI species with a stronger transition can serve as an optical interface for high-fidelity readout and entanglement protocols \cite{walther_high-fidelity_2015,asadi_quantum_2018}. Overall, this offers a promising potential for scalable quantum processing nodes \cite{kinos_high-connectivity_2022,kinos_roadmap_2021}.\\ 

Despite these favorable properties, progress toward single-ion experiments and scalable architectures with \ce{Eu^{3+}} has been hindered by fundamental limitations. In particular, the long excited state lifetime and the weak branching ratio of the coherent $\mathrm{}^5\mathrm{D}_0\rightarrow \mathrm{}^7\mathrm{F}_0$ transition, typically below \SI{1}{\percent}, result in a low photon emission rate, rendering efficient optical readout experimentally challenging. Although significant efforts have been devoted to enhancing the emission by integrating \ce{Eu^{3+}}-doped crystals into optical microcavities \cite{Casabone_2018,Eichhorn}, optical detection of single \ce{Eu^{3+}} ions has not yet been achieved. Furthermore, the random doping of \ce{Eu^{3+}} in solid state hosts leads to uncontrolled and heterogeneous ion-ion coupling strength. These persistent drawbacks highlight the need for approaches that increase the branching ratio of the coherent transition while preserving the exceptional coherence properties of \ce{Eu^{3+}}-based systems, and enable the design of defined multi-qubit registers.\\

An emerging approach to overcome these limitations is to incorporate REI into molecular matrices \cite{bayliss_optically_2020,kuppusamy_spin-bearing_2024}. In such systems, the ligand coordination can be chemically engineered with atomic precision, offering new degrees of freedom to tailor branching ratios, ion-ion separations, and the local spin environment. Recent studies on REI-based molecular complexes have demonstrated promising optical \cite{serrano_ultra_2022} and spin \cite{weiss_high-resolution_2025,Vasilenko2026} coherence properties at cryogenic temperatures, indicating that the favorable coherence characteristics of REI can be preserved outside conventional crystalline hosts. At the same time, the molecular approach opens a pathway toward deterministic multi-ion architectures within a single molecule, providing a natural platform for optically controlled ion-ion interactions and elementary multi-qubit gate operations.

\section{Results and Discussion}
\label{results_discussion}

In this work, we investigate the optical properties of two \ce{Eu^{3+}}-based complexes in order to assess the suitability of molecular systems as optically addressable spin qubits for quantum information applications. We study a mononuclear complex as a reference platform \cite{Batista} and a dinuclear analogue \cite{Ilmi}, in which the position and distance of two \ce{Eu^{3+}} centers is well-defined. Such chemically engineered multi-core architectures can enable spectrally selective optical addressing of individual ions within a single molecule and allow electric dipole-dipole interactions between them to be probed. We observe excellent optical coherence properties and find clear signatures of nuclearity-dependent ion-ion interactions in the mono- and dinuclear complexes, laying the grounds for controlled  two-qubit gate operations. We furthermore integrate the dinuclear complex into a fiber-based open-access microcavity and observe Purcell-enhanced emission, boosting both the radiative quantum yield and the branching ratio of the coherent $\mathrm{}^5\mathrm{D}_0\rightarrow\mathrm{}^7\mathrm{F}_0$ transition to large values. This underlines the potential of molecular REI systems as a viable qubit platform for scalable quantum technologies.\linebreak

\begin{figure}[tb] 
\centerline{\includegraphics[width=\columnwidth]{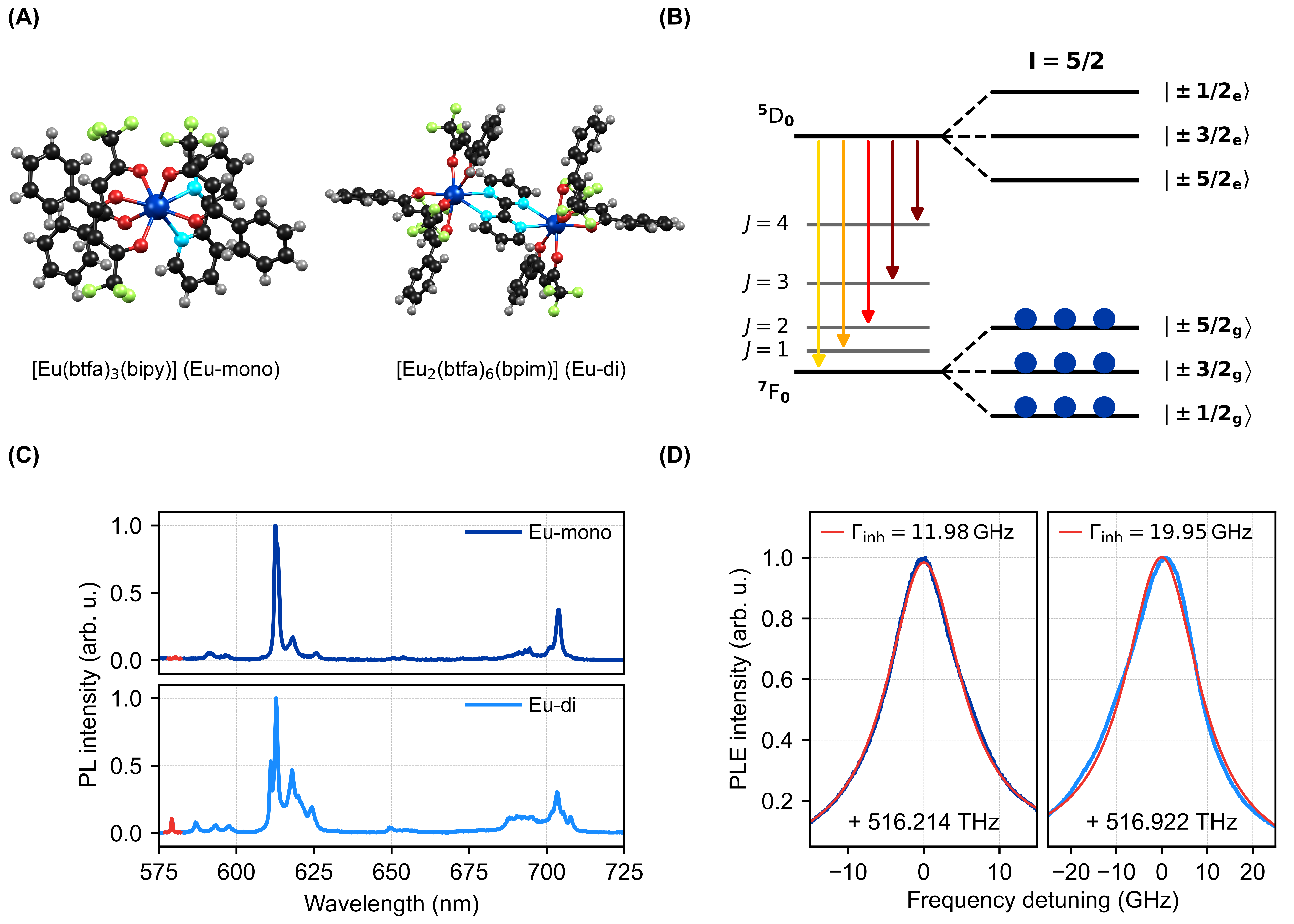}}
\caption{\textbf{Photoluminescence properties of the mononuclear and dinuclear Eu\textsuperscript{3+} complexes. (A):} Molecular structures of Eu-mono (left) and Eu-di (right). \textbf{(B):} Energy-level scheme of \ce{Eu^{3+}} showing the relevant radiative decay channels from the $\mathrm{}^5\mathrm{D}_0$ excited state into the $\mathrm{}^7\mathrm{F}_J$ manifold as observed in the emission spectra. \textbf{(C):} Room-temperature photoluminescence spectra of Eu-mono (upper panel) and Eu-di (lower panel). The dinuclear complex shows a significantly higher branching ratio of the coherent $\mathrm{}^5\mathrm{D}_0 \rightarrow\mathrm{}^7\mathrm{F}_0$ transition, which is highlighted in red in the spectra. \textbf{(D):} Photoluminescence excitation spectra recorded at about \SI{100}{\milli\kelvin}, revealing the inhomogeneous broadening of the $\mathrm{}^5\mathrm{D}_0 \rightarrow \mathrm{}^7\mathrm{F}_0$ transition. Lorentzian fits yield linewidths of \SI{11.98 \pm 0.05} for Eu-mono (left panel) and \SI{19.95 \pm 0.04}{\giga\hertz} for Eu-di (right panel).} 
\label{fig:F1}
\end{figure}

The chemical structures of the mononuclear complex \ce{[Eu(btfa)3(bipy)]} (Eu-mono) and the dinuclear complex \ce{[Eu2(btfa)6(bpim)]} (Eu-di) are shown in Figure~\ref{fig:F1}(A). In the asymmetric unit of both complexes, the \ce{Eu^{3+}} center is coordinated by three 4,4,4-trifluoro-1-phenyl-1,3-butanedione (btfa) ligands and two nitrogen donors from the bidentate ligand 2,2'-bipyridine (bipy) in Eu-mono. In the dinuclear complex, two such \ce{Eu^{3+}} coordination units are linked by a 2,2'-bipyrimidine (bpim) bridging ligand, yielding a molecule with two optically equivalent centers at a well-defined distance of about \SI{7}{\angstrom}. A detailed description of the chemical synthesis and crystal structures of both complexes can be found in the Supporting Information. All measurements reported in the following were conducted on microcrystalline samples with grain sizes up to \SI{50}{\micro\meter} and a \ce{Eu^{3+}} doping concentration of \SI{100}{\percent}. The samples contain a natural abundance mixture of \ce{^{151}Eu} and \ce{^{153}Eu} isotopes.\linebreak

\subsection{Photoluminescence Spectra}

The photoluminescence (PL) properties of both molecular complexes were characterized at room temperature under off-resonant excitation at \SI{532}{\nano\meter} using a confocal microscope, and at cryogenic temperatures under ultraviolet (UV) excitation at \SI{365}{\nano\meter} using a commercial spectrofluorometer. The emitted fluorescence was collected after spectral filtering to suppress the excitation light. Figure~\ref{fig:F1}(B) schematically illustrates the characteristic intra-$4f$ transitions of \ce{Eu^{3+}} from the lowest excited $\mathrm{}^5\mathrm{D}_0$ state to the different $J$-levels of the $\mathrm{}^7\mathrm{F}_J$ ground state. Each individual radiative decay channel exhibits a distinct branching ratio. These differences are directly reflected in the recorded intensities of the emission lines in the PL spectra up to the $J$=4 value, as shown in Figure~\ref{fig:F1}(C). Excitation and emission spectra measured at \SI{3}{\kelvin}, together with their detailed analysis, are provided in the Supporting Information (Section~2).\linebreak

Although both complexes are structurally assigned to a $D_{4d}$ symmetry, the clear observation of the coherent electric dipole $\mathrm{}^5\mathrm{D}_0\rightarrow\mathrm{}^7\mathrm{F}_0$ transition indicates a deviation from the structural site symmetry \cite{Senthil_2026_Symmetry}. As discussed by Binnemans \cite{Binnemans}, \ce{Eu^{3+}}-based materials require point group symmetries such as $C_n$, $C_{nv}$, or $C_s$ to partially relax the selection rules and thus induce the $\mathrm{}^5\mathrm{D}_0\rightarrow\mathrm{}^7\mathrm{F}_0$ transition. The observed branching ratio of the $\mathrm{}^5\mathrm{D}_0\rightarrow\mathrm{}^7\mathrm{F}_0$ for Eu-di is \SI{1.30}{\percent}, which is significantly higher than the \SI{0,14}{\percent} observed for the mononuclear analogue. Such an enhanced intensity of this transition is consistent with previously reported dinuclear complexes \cite{Baker2009,Errulat2018} featuring similar structural motifs, where inter-lanthanide dipolar interactions may provide an additional perturbation that facilitates $J$-mixing. The corresponding $\mathrm{}^5\mathrm{D}_0\rightarrow\mathrm{}^7\mathrm{F}_0$ emission is centered at \SI{580,548}{\nano\meter} with an FWHM of \SI{0.154 \pm 0.010}{\nano\meter} for the mononuclear complex and at \SI{579,627}{\nano\meter} (both values in vacuum) with a fitted FWHM of \SI{0.340 \pm 0.002}{\nano\meter} for the dinuclear complex.

\subsection{Cryogenic Ensemble Spectroscopy}

\begin{figure}[tb] 
\centerline{\includegraphics[width=\columnwidth]{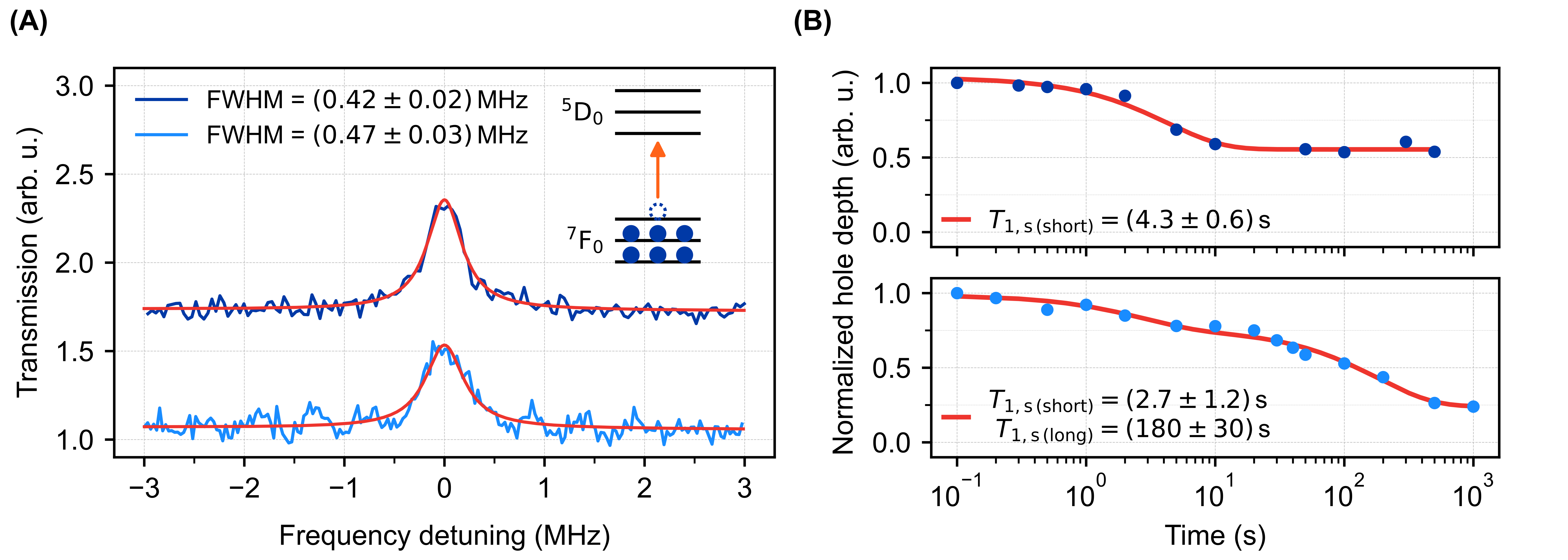}}
\caption{\textbf{Spectral hole burning and hyperfine state lifetime. (A):} Absorption-based spectral hole burning measurements for both complexes, shown with a vertical offset for clarity. Lorentzian fits yield full width at half maximum (FWHM) linewidths of \SI{0.42\pm0.02}{\mega\hertz} for Eu-mono and \SI{0.47\pm0.03}{\mega\hertz} for Eu-di. \textbf{(B):} Hole depth evolution for both complexes. Eu-mono (upper panel) exhibits a single-exponential decay with a relaxation time of \SI{4.3\pm 0.6}{\second}, and no further decay within the \SI{500}{\second} measurement window, suggesting hour-long spin lifetime. Eu-di (lower panel) shows a bi-exponential decay with time constants of \SI{2.7\pm1.2}{\second} and \SI{180\pm30}{\second}.} 
\label{fig:F2}
\end{figure}

The samples were integrated into a custom-built sample holder, which was mounted on the cold plate of a dilution refrigerator. Measurements were predominantly performed at temperatures of approximately \SI{100}{\milli\kelvin}. Optical excitation and detection were performed via a fiber-based coupling scheme, enabling a compact cryogenic setup.\\
As a first step, a tunable dye laser was scanned across the $\mathrm{}^7\mathrm{F}_0\rightarrow\mathrm{}^5\mathrm{D}_0$ transition to probe the optical inhomogeneous line $\Gamma_{\text{inh}}$ of the \ce{Eu^{3+}} ensembles. The resulting photoluminescence excitation (PLE) spectra for the two complexes are shown in Figure~\ref{fig:F1}(D). Lorentzian fits yield narrow inhomogeneous linewidths of \SI{11,05\pm 0,05}{\giga\hertz} for Eu-mono and \SI{19,95\pm 0,04}{\giga\hertz} for Eu-di. Such linewidths are characteristic of previously studied REI-based powder materials, in which structural disorder is more pronounced and therefore leads to still broader inhomogeneous profiles compared to highly ordered macroscopic single crystals \cite{Vasilenko2026}.\\
To access the homogeneous linewidth $\Gamma_{\text{h}}$ and probe the hyperfine spin states, spectral hole burning (SHB) measurements were performed. To infer the spin lifetime $T_{1,\text{s}}$, spectral holes were burned at an optical power of approximately \SI{1}{\milli\watt}, providing sufficient hole contrast while limiting the thermal load on the sample. For probing, a significantly lower excitation power of about \SI{15}{\micro\watt} was used to minimize power broadening. Under these conditions, narrow spectral holes were recorded for both Eu-mono and Eu-di, as shown in Figure~\ref{fig:F2}(A). The extracted homogeneous linewidth yields $\Gamma_{\text{h}}=\SI{210}{\kilo\hertz}$ for Eu-mono and $\Gamma_{\text{h}}=\SI{235}{\kilo\hertz}$ for Eu-di, corresponding to an optical coherence time $T^*_{2,\text{o}}=1/(\pi\Gamma_{\text{h}})=\SI{1,5}{\micro\second}$ and \SI{1,3}{\micro\second}, respectively. We note that this approaches the linewidth of the laser and is thus an upper bound. These exceptionally narrow hole widths indicate excellent optical coherence maintained over long timescales and underscore the potential of these molecular systems as candidates for optically addressable qubits.\\ 

The hyperfine spin lifetime was assessed by monitoring the decay of the spectral hole depth as a function of the delay time between the burn and probe pulses, as shown in Figure~\ref{fig:F2}(B). For Eu-mono (upper panel), the hole depth follows a single-exponential decay with a relaxation time of $T_{1,\text{s}}=\SI{4,3\pm0,6}{\second}$ until the normalized hole depth approaches \SI{\sim0.5}{}. Remarkably, no further decay is observed within the \SI{500}{\second} measurement window, indicating hour-long spin lifetime of a sub-population.\\

In contrast, Eu-di (lower panel) exhibits a clearly distinguishable bi-exponential decay, characterized by relaxation times of $T_{1,\text{s(short)}}=\SI{2,7\pm1,2}{\second}$ and $T_{1,\text{s(long)}}=\SI{180\pm30}{\second}$. The origin of this multi-component behavior is not yet fully understood and may arise from several mechanisms, including phonon-mediated processes and coupling to paramagnetic impurities. Moreover, the dinuclear bridging structure may introduce additional interactions that affect the spin relaxation dynamics. Notably, the observed spin state lifetimes are comparable to those reported for REI-based solid-state platforms \cite{kuppusamy_observation_2023,Schlittenhardt,Liu2025,Konz_2003}.\linebreak

\begin{figure}[tb] 
\centerline{\includegraphics[width=\columnwidth]{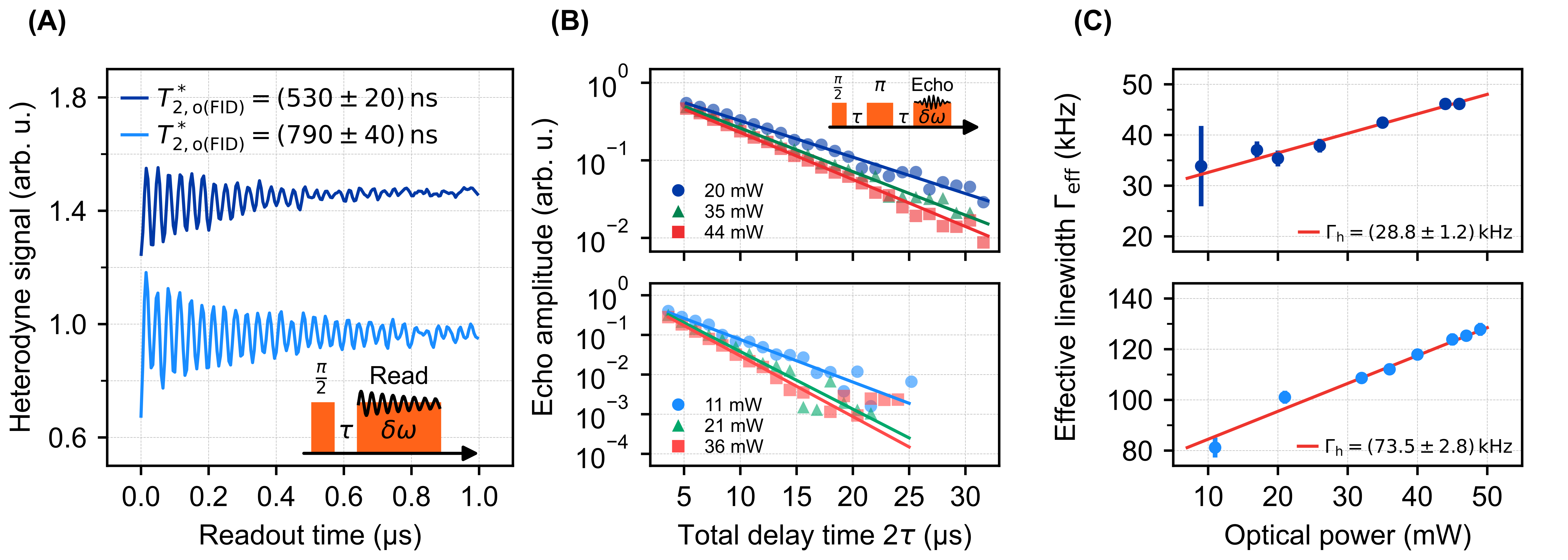}}
\caption{\textbf{Optical coherence measurements. (A):} Optical free-induction decay measurements obtained using a $\pi/2$ pulse followed by a heterodyne readout pulse. The two traces are vertically offset for clarity and yield optical dephasing times of $T^*_{2,\text{o(FID)}}=\SI{530\pm20}{\nano\second}$ for Eu-mono and $T^*_{2,\text{o(FID)}}=\SI{790\pm40}{\nano\second}$ for Eu-di. \textbf{(B):} Two-pulse photon echo decays recorded at different excitation powers, with three power levels shown in each panel. Increasing optical power accelerates the echo decay and reduces the optical coherence time $T_{2,\text{o}}$. The longest observed echo decays are approximately \SI{9}{\micro\second} for Eu-mono and \SI{4}{\micro\second} for Eu-di. \textbf{(C):} Effective homogeneous linewidth $\Gamma_{\text{eff}}$ as a function of excitation power for the two complexes, extracted from the photon echo measurements. Extrapolation to zero power yields the intrinsic homogeneous linewidth $\Gamma_{\text{h}}$.}
\label{fig:F3}
\end{figure}

While SHB provides an upper bound for the homogeneous linewidth, it does not directly probe the coherence of the optical transitions. To better access the optical coherence properties and dephasing dynamics of the \ce{Eu^{3+}} ensembles, free-induction decay (FID) measurements were therefore performed for both Eu-mono and Eu-di. The resulting FID traces are shown in Figure~\ref{fig:F3}(A), revealing a lower bound of the instantaneous pure dephasing time $T^*_{2,\text{o(FID)}}$ in both molecular systems. The FID signals were recorded using a pulsed excitation scheme in which an optical $\pi/2$ pulse prepares an optical coherence, followed by heterodyne detection of the coherent emission signal using a frequency-shifted readout pulse. From fits to the decay of the oscillatory beating signal, pure dephasing times of $T^*_{2,\text{o(FID)}}=\SI{530\pm20}{\nano\second}$ for Eu-mono and $T^*_{2,\text{o(FID)}}=\SI{790\pm40}{\nano\second}$ for Eu-di were extracted. The shorter $T^*_{2,\text{o(FID)}}$ times compared to those found in SHB indicate that power-induced effects play a significant role. In particular, mechanisms such as power broadening and instantaneous spectral diffusion (ISD) are expected to contribute to the observed dephasing, in accordance with the higher power levels used for the FID experiment (Experimental Section~\ref{sec:exp_section}).\linebreak

To obtain a quantitative measure of the intrinsic optical coherence time $T_{2,\text{o}}$, two-pulse photon echo measurements were subsequently performed using the same heterodyne detection scheme as employed for the FID measurements. Figure~\ref{fig:F3}(B) shows the two-pulse photon echo decays for Eu-mono (upper panel) and Eu-di (lower panel) recorded at three different optical excitation powers. In both molecular systems, the photon echo amplitude decays exponentially with increasing delay times $\tau$. However, a pronounced difference is observed between the two molecules: at the lowest excitation power, Eu-mono exhibits a longer optical coherence time of $T_{2,\text{o}}\approx\SI{9}{\micro\second}$, whereas Eu-di shows a shorter coherence time of $T_{2,\text{o}}\approx\SI{4}{\micro\second}$. For both complexes, increasing the optical excitation power leads to a systematic reduction of $T_{2,\text{o}}$, indicating the presence of excitation-induced decoherence mechanisms. The more pronounced loss of coherence observed for the dinuclear complex highlights the impact of the molecular architecture on optical dephasing, likely arising from increased ISD within the dinuclear structure.\\

To probe excitation-induced dephasing more quantitatively, power-dependent photon echo measurements were performed and analyzed to extrapolate to the homogeneous linewidth in the limit of vanishing power. The corresponding effective linewidths were extracted from each photon echo decay according to $\Gamma_{\text{eff}}=1/(\pi T_{2,\text{o}})$, and are summarized in Figure~\ref{fig:F3}(C). For both Eu-mono (upper panel) and Eu-di (lower panel), $\Gamma_{\text{eff}}$ increases linearly with optical excitation power. The power dependence associated with ISD was analyzed using the model described in Ref.~\cite{Thiel2014}, which is valid in the weak-excitation regime. Within this framework, the effective homogeneous linewidth is given by:
\begin{equation}
    \lim_{P\rightarrow0}\Gamma_{\text{eff}}=\Gamma_{\text{h}}+\frac{1}{2}\gamma_{\text{ISD}}P,
\end{equation}
where $\gamma_{\text{ISD}}$ denotes the linear ISD coefficient and $\Gamma_{\text{h}}$ the intrinsic homogeneous linewidth extrapolated to zero excitation power. Across the investigated power range, Eu-mono consistently exhibits a narrower effective homogeneous linewidth than Eu-di, with extrapolated zero-power values of $\Gamma_{\text{h}}=\SI{28.8\pm1.2}{\kilo\hertz}$ for Eu-mono and \SI{73,5\pm2,8}{\kilo\hertz} for Eu-di and slopes of \SI{0,77\pm0,06}{\kilo\hertz\per\milli\watt} and \SI{2,20\pm0,15}{\kilo\hertz\per\milli\watt}, respectively.\\

These results confirm that excitation-induced dephasing constitutes a significant broadening mechanism, while interactions between the optical centers within the dinuclear molecular complex contribute to further broadening.

\subsection{Optically Controlled Ion-Ion Interactions}

\begin{figure}[tb] 
\centerline{\includegraphics[width=\columnwidth]{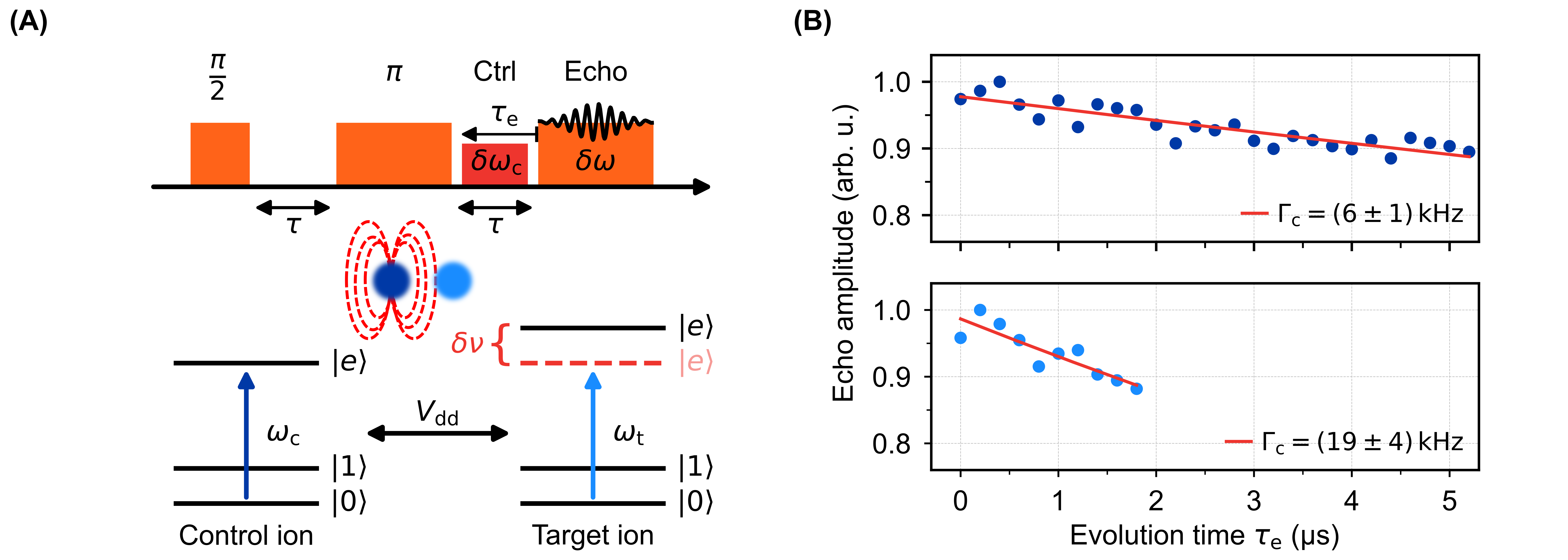}}
\caption{\textbf{Probing ion-ion interactions via controlled excitation of a sub-ensemble. (A):} Pulse sequence used to probe ion-ion interactions (top panel) and schematic illustration of the control-target ion configuration (bottom panel). A frequency-selective control pulse (red pulse) is applied to a subset of ions at frequency $\omega_{\text{c}}$, while the optical coherence of spectrally distinct target ions at $\omega_{\text{t}}$ is monitored using a two-pulse photon echo sequence (orange pulses). The control and target ions are represented as effective three-level systems with ground states $\ket{0}, \ket{1}$ and optically excited state $\ket{e}$. The interaction between control and target ions is described by the dipole--dipole coupling strength $V_{\text{dd}}$, which induces a frequency shift $\delta\nu$ of the excited state of the target ions. \textbf{(B):} Photon echo amplitude of the target ions as a function of the evolution time $\tau_{\text{e}}$ between the control pulse and the photon echo readout, shown for Eu-mono (upper panel) and Eu-di (lower panel). The decay of the echo amplitude reflects interaction-induced dephasing of the target ions caused by the excitation of the control ions. Solid red lines are exponential fits, from which an effective interaction-induced broadening $\Gamma_{\text{c}}$ of the homogeneous linewidth of the target ions is extracted.}
\label{fig:F4}
\end{figure}

Realizing optically controlled interactions between individual ions is a central requirement for implementing optical qubit gate operations, multi-qubit control, and readout protocols in REI-based quantum platforms. As a first step toward a two-qubit gate operation, we investigate optically induced ion-ion interactions at the ensemble level in our molecular systems, similar to Ref.~\cite{serrano_ultra_2022}.
Controlled interaction between \ce{Eu^{3+}} ions is implemented using the pulse sequence shown in Figure~\ref{fig:F4}(A) (upper panel), which selectively excites a spectrally detuned control sub-ensemble while monitoring the optical coherence of target ions with a photon echo sequence.\\

Optical excitation of the control ions changes their permanent electric dipole moment in the excited state, resulting in a change of the local electric field experienced by nearby target ions, as schematically illustrated in Figure~\ref{fig:F4}(A) (lower panel). This interaction leads to a dipole-dipole-mediated shift of the optical transition frequency of the target ions, resulting in an additional broadening of the homogeneous linewidth. Control and target ions are addressed at different frequencies $\omega_{\text{c}}$ and $\omega_{\text{t}}$, respectively, within the inhomogeneous absorption profile, exploiting the concept of spectrally distributed ion classes for selective optical addressing. The characteristic Eu--Eu distances differ between the two systems: in the mononuclear complex, the nearest separation is intermolecular ($\SI{\approx8,5}{\angstrom}$), whereas in the dinuclear complex, an intramolecular distance of \SI{\approx7}{\angstrom} is present, with the nearest intermolecular separation being \SI{\approx9,8}{\angstrom}. These distances define the strength of the dipole--dipole interaction and are discussed in detail in the Supporting Information (Section~4). To experimentally quantify the strength of the ion--ion interaction, we monitor the photon echo amplitude of the target ions as a function of the evolution time $\tau_{\text{e}}$, defined as the delay between the onset of the control pulse and the photon echo readout. In this sequence, the control pulse is applied between the optical $\pi$ and readout pulses of the photon echo, as shown in Figure~\ref{fig:F4}(A) (upper panel). The resulting decay of the echo amplitude reflects an effective interaction-induced broadening $\Gamma_{\text{c}}$ of the homogeneous linewidth, arising from the distribution of ion-ion interaction strengths within the excited control sub-ensemble.\\

Exponential fits of the form $\exp(-\pi\Gamma_{\text{c}}\tau_{\text{e}})$ to the data in Figure~\ref{fig:F4}(B) yield interaction-induced broadenings of $\Gamma_{\text{c}}=\SI{6\pm1}{\kilo\hertz}$ for Eu-mono (upper panel) and \SI{19\pm4}{\kilo\hertz} for Eu-di (lower panel). These results provide direct evidence of optically controlled ion-ion interactions in both molecular systems, with the dinuclear complex exhibiting stronger interaction-induced dephasing. In the current experiments, the ions that contribute to the interaction are still randomly distributed with a relatively low density (Supporting Information, Section~4). With more advanced ion class distillation \cite{longdell_demonstration_2004}, we expect that it will be possible to select subsets of ion pairs with strong coupling, such that intramolecular pairs, as found in the dinuclear complex, exhibiting a deterministic interaction strength, can be specifically targeted. This demonstrates that conditional control is feasible using the ensemble approach in molecular REI-based materials and highlights multinuclear molecular architectures as a promising resource for realizing optically mediated two- and multi-qubit gate operations.

\subsection{Purcell Enhancement of Optical Emission}

A remaining challenge is the integration of such molecular systems into photonic environments to enhance light-matter interactions. In the following, we address this challenge by integrating the dinuclear complex into a cryo-compatible, fully-tunable, open-access, fiber-based Fabry-P\'erot microcavity as detailed in previous works \cite{Casabone_2018,Pallmann, Eichhorn}.\\ 
Coupling to micro- and nano-cavities has been used to enhance the emission rate of various emitters, including REIs, such as: \ce{Eu^{3+}}:\ce{Y2O3} \cite{Eichhorn, Casabone_2018, Sinan_2013}, \ce{Er^{3+}}:\ce{Y2O3} \cite{Deshmukh:23}, \ce{Er^{3+}}:\ce{Y2SiO5} \cite{Raha_2020, Ulanowski22_SciAdv}, and \ce{Yb^{3+}}:\ce{YVO4} \cite{Kindem_2020,ruskuc_multiplexed_2025}. This has enabled single ion readout and control \cite{kindem_control_2020}, spectral multiplexing \cite{chen_parallel_2020,Ulanowski22_SciAdv}, and entanglement distribution \cite{ruskuc_multiplexed_2025}. Recently, cavity-enhanced emission has been observed in a molecular \ce{Eu^{3+}} complex coupled to a planar Fabry-P\'erot cavity \cite{emmanuele_microcavity_2022} and to a nanobeam cavity \cite{emmanuele_amplified_2023}.\\
For the case of the stable Fabry-P\'erot cavity used here, the enhancement of the emission is quantified by the Purcell factor defined in the following way:
\begin{equation}
F_{\text{P}} = \frac{6}{\pi^{3}}\bigg(\frac{\lambda}{n}\bigg)^{2}\frac{\mathcal{F}}{w_{0}^{2}} \eta_E \eta_t,
\end{equation}
where $\lambda$ is the wavelength of the transition of interest, $n$ the refractive index of the thin-film, $\mathcal{F}$ is the finesse of the cavity, $w_{0}$ is the waist of the Gaussian cavity mode, and $\eta_E=\left(\vec{d}\cdot \vec{E}(r_0)/(d\cdot E_\mathrm{max})\right)^2 $ describes the dipole overlap with the dipole moment $d$, the electric field $E$, the location of the ion $r_0$, and the maximum of the standing wave field in the cavity $E_\mathrm{max}$. Finally, $\eta_t$ describes a correction factor that accounts for the cavity electric field distribution in the presence of the thin-film \cite{janitz_fabry-perot_2015,van_dam_optimal_2018}.\linebreak

\begin{figure}[tb] 
\centerline{\includegraphics[width=\columnwidth]{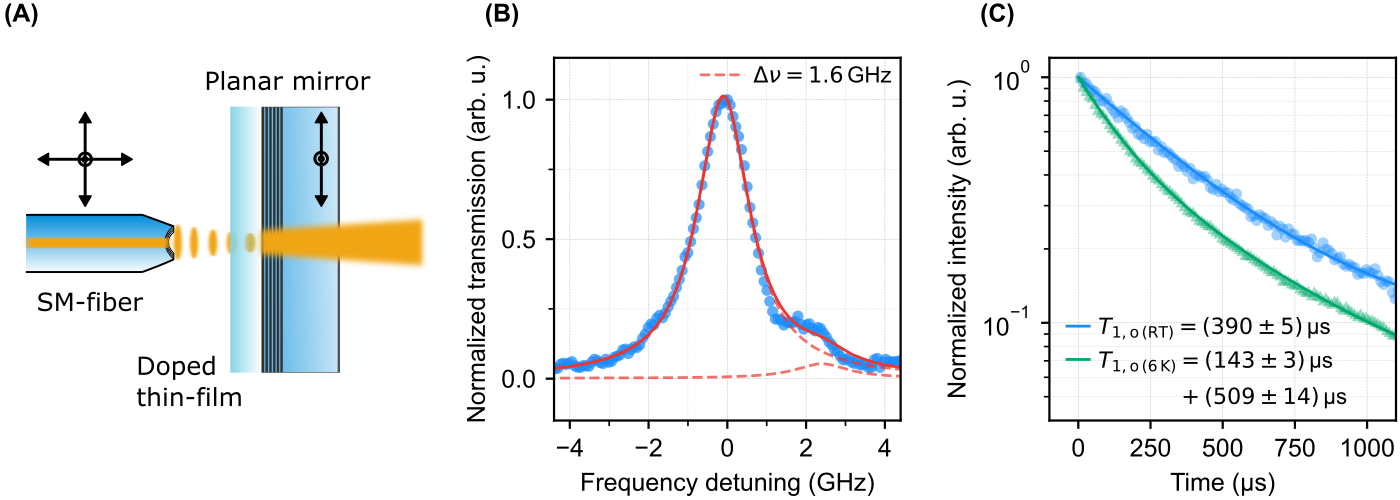}}
\caption{\textbf{Purcell enhancement of the coherent transition. (A):} Schematic illustration of the fiber-based microcavity setup employed in this work. \textbf{(B):} Fitted cavity resonance in the presence of the thin-film doped with Eu-di. The second polarization mode visible at a frequency detuning of \SI{2.4}{\giga\hertz} arises from the slight asymmetry in the concave profile of the fiber. \textbf{(C):} Purcell enhancement of the coherent $\mathrm{}^5\mathrm{D}_0\rightarrow\mathrm{}^7\mathrm{F}_0$ transition at room temperature (blue), and at \SI{6}{\kelvin} (green).} 
\label{fig:F5}
\end{figure}

A schematic of the cavity is depicted in Figure~\ref{fig:F5}(A). We prepare a poly(methyl methacrylate) (PMMA) thin-film ($n=1.49$) doped with Eu-di on a planar mirror and form a tunable microcavity with a second micro-mirror fabricated on the end facet of an optical fiber. Information pertaining to the cavity performance and thin-film preparation is given in the experimental methods. Figure~\ref{fig:F5}(B) depicts a cavity resonance measured by sweeping the cavity length at a fixed probe wavelength of \SI{580.04}{\nano\meter}, resulting in a cavity linewidth of \SI{1.6}{\giga\hertz} when converting the length change into a frequency, and a corresponding quality factor $Q=\SI{3e5}{}$. We also measure the cavity finesse and find a value of $\mathcal{F}=21{,}500$, which is larger than for the cavity without the thin-film (Experimental Section~\ref{sec:exp_section}), indicating that we are operating approximately under air-like mode conditions \cite{janitz_fabry-perot_2015,van_dam_optimal_2018}, where $\eta_d\approx 1/n$. Together with the inferred mode waist $w_0=\SI{1.1}{\micro\meter}$, we can calculate the expected maximal Purcell factor ($\eta_E=1$) to be $F_\text{P}=\SI{350}{}$.\\

We measure the emission lifetime of the ions under resonant coupling of a cavity resonance with the $\mathrm{}^5\mathrm{D}_0\rightarrow\mathrm{}^7\mathrm{F}_0$ transition both at room temperature and at \SI{6}{\kelvin}, as depicted in Figure ~\ref{fig:F5}(C). At room temperature, we observe a single exponential with a lifetime of \SI{390 \pm 5}{\micro\second}. We note that while this lifetime was measured in the cavity, we estimate that, due to the broad homogeneous linewidth at room temperature, the observed lifetime matches that of the free-space, which was confirmed by confocal measurements on the same sample. At \SI{6}{\kelvin}, the observed lifetime becomes non-exponential, and we performed a double exponential fit resulting in two time constants of \SI{143 \pm 3}{\micro\second} and \SI{509\pm 14}{\micro\second}. The deviation from a single-exponential is expected due to the distribution of dipole orientations and locations within the cavity field, and time-dependent cavity length jitter, resulting in a distribution of Purcell enhancements. Here, the shorter lifetime component evidences Purcell enhancement for a subset of ions with a favorable dipole-alignment and location within the cavity field ($\eta_E\rightarrow 1$), whereas the long time component represents ions that are not well coupled to the cavity field ($\eta_E\rightarrow 0$) and therefore approximates the free-space lifetime. It is noted that the long time component is longer than the lifetime measured at room temperature. We attribute this to a possible improvement of the quantum efficiency $\mathrm{QE}$ at low temperature, and to a lateral drift of the cavity upon cooldown, resulting in the lifetime measurement occurring at a location with a different local environment and thin-film thickness which can affect the radiative lifetime. From the two lifetime components measured at \SI{6}{\kelvin}, we calculate the effective Purcell factor at this position in the thin-film, defined as:
\begin{equation}
F_{\text{P}}^{\text{eff}} = \zeta \cdot \mathrm{QE} \cdot F_{\text{P}} = \frac{T_{1,\text{o(FS)}}}{T_{1,\text{o(cav)}}} - 1.
\end{equation}
Here, $T_{1,\text{o(FS)}}$ and $T_{1,\text{o(cav)}}$ denote the free-space and cavity-enhanced optical lifetimes, respectively. The parameter $\zeta$ is the branching ratio of the $\mathrm{}^5\mathrm{D}_0\rightarrow\mathrm{}^7\mathrm{F}_0$ transition, calculated to be \SI{1.4}{\percent} \cite{Ilmi}. The quantum efficiency $\mathrm{QE}$ for our sample is determined to be \SI{\sim 0.480}{} using the radiative lifetime reported in Ref.~\cite{Ilmi}. Details of the quantum efficiency calculation are provided in the Supporting Information (Section~3). From the lifetime change, we obtain an effective Purcell factor of \SI{2.55 \pm 0.25}{}, and with $\zeta$ and $\mathrm{QE}$ as given above, we infer an ideal Purcell factor of $F_P=\SI{380 \pm 40}{}$. This value directly describes the enhancement of the $\mathrm{}^5\mathrm{D}_0\rightarrow\mathrm{}^7\mathrm{F}_0$ transition and is in good agreement with the estimated maximum Purcell factor calculated above.\linebreak

We note that while the cavity-induced lifetime reduction is moderate, the branching ratio and quantum efficiency of the system are strongly increased. Cavity-enhancement increases the fraction of excitations ending up in the emission of $\mathrm{}^5\mathrm{D}_0\rightarrow\mathrm{}^7\mathrm{F}_0$ photons from the free-space value of $\eta_0=\zeta\cdot\mathrm{QE}=\SI{6,7e-3}{}$ by 108-fold to $\eta_C=(F_P+1)/(F_P+(\zeta\cdot\mathrm{QE})^{-1})=\SI{0.72}{}$. Furthermore, the emission is coupled into a well-collectable cavity mode with an efficiency of $\beta=F_{\text{P}}^{\text{eff}} /(F_{\text{P}}^{\text{eff}} +1)=\SI{0.72}{}$, such that \SI{35}{\percent} of optical excitations lead to photons emitted into the cavity mode, from which a large fraction (\SI{\sim65}{\percent}) can be outcoupled and detected.

\section{Conclusion}
\label{sec:conclusion}

In summary, we have demonstrated that \ce{Eu^{3+}}-based molecular complexes provide a versatile and chemically tunable platform with the potential to design multi-qubit architectures that enable optical control of electric dipole-dipole interactions relevant for two-qubit gate operations. We systematically compare mono- and dinuclear \ce{Eu^{3+}} complexes from the same molecular family, and observe long optical coherence times and long spin lifetimes. We identify that the optical power-dependent broadening is stronger in the dinuclear complex, which indicates the presence of ion-ion interactions. We controllably harness ion-ion interactions with a control-target protocol to show the basic mechanism required for the realization of two-qubit quantum gates, and find a more than three-fold larger interaction-induced dephasing rate for the dinuclear vs mononuclear complex. This highlights multinuclear molecular design as a deterministic, chemistry-enabled concept for realizing optically mediated two- and multi-qubit gate schemes in REI-based quantum platforms. At the same time, excitation-induced dephasing and interaction-driven broadening define an important optimization landscape, motivating future studies to explore reduced dopant concentrations and more complex multinuclear architectures with tailored intramolecular ion-ion separations to balance the preservation of optical coherence and optically controlled interaction strength. Finally, we demonstrate that integration of molecular complexes into an optical microcavity enables Purcell-enhanced emission and a strong increase in quantum yield and branching ratio. Further progress will rely on improved sample transfer and recrystallization techniques that preserve the intrinsic optical properties of the molecular complexes, paving the way toward scalable molecular quantum technologies that bridge well-established properties of REI in solid-state host materials with the tunability enabled by molecular chemical engineering.

\newpage

\section{Experimental Section}
\label{sec:exp_section}

\hspace*{1.5em}\textit{Materials:} \ce{EuCl3.6H2O}, 4,4,4-trifluoro-1-phenyl-1,3-butanedione, 2,2’-bipyridine and 2,2’-bipyrimidine were purchased from commercially available sources and used without further purification. Ethanol (\SI{96}{\percent}) and distilled water were used for the synthesis and crystallization.\\

\par\noindent\hspace*{1.5em}\textit{Synthesis and Characterization:} The modified synthesis of the mononuclear and dinuclear complexes compared with published procedures \cite{Batista, Ilmi} is described in the Supporting Information (Section~1). The complexes can be synthesized either through a one-pot reaction, where all reactants are combined in a single vessel without isolating intermediates, or via a two-step synthetic path. In the latter approach, when a lanthanide salt reacts with a selected $\beta$-diketonate ligand in a 1:3 molar ratio in aqueous solution, an aqua mononuclear complex is formed and can be isolated in the first step. In order to avoid O--H vibrations from coordinated water molecules, which often reduce the PL properties of lanthanide complexes due to enhanced non-radiative decay processes, it is more efficient to replace these ligands. This can be achieved in a second step, where the mononuclear precursor reacts with the corresponding N-donor heterocyclic ligand. The advantage of the second approach is the high solubility of both reactants in small solvent volumes, which enables the formation high-quality crystals within several days.\\
All products were characterized by elemental analysis, powder X-ray diffraction (P-XRD), and single-crystal X-ray diffraction (SC-XRD) techniques. The unit cell parameters were determined using a STOE IPDS 2T diffractometer with an image plate detector and \ce{Mo} K$\alpha$ radiation. Powder diffraction patterns were collected on a STOE Stadi P diffractometer at room temperature using \ce{Cu} K$\alpha$ radiation. Simulated P-XRD patterns were obtained from CIF files (CCDC deposition numbers: 1176296 (Eu-mono), 2000896 (Eu-di), and 1273298 (precursor)) using the Mercury program \cite{macrae2020mercury}. PL measurements at \SI{3}{\kelvin} were performed on a Horiba Fluorolog spectrometer with a 920 photomultiplier tube detector. The crystalline samples were placed between two quartz glass plates with a drop of perfluorinated oil. A \SI{400}{\nano\meter} longpass filter was used to suppress second- and higher-order diffraction peaks in the spectra.\\

\par\noindent\hspace*{1.5em}\textit{Fiber-Based Ferrule Sample Holder:} The samples are incorporated into a home-built sample holder consisting of two opposing cylindrical stainless steel ferrules with a diameter of \SI{2,5}{\milli\meter}, connected by a phosphor bronze mating sleeve. Multimode optical fibers with a core diameter of \SI{200}{\micro\meter} are integrated into the ferrules using a cryo-compatible UV-curing adhesive. One of the fibers is retracted by approximately \SI{200}{\micro\meter}, thereby defining the sample space between the fiber facets. In this configuration, one fiber is used for excitation, while the second fiber collects the transmitted signal.\\
After sample loading, the ferrules are inserted into the mating sleeve and fixed in place using the same UV-curing adhesive to ensure mechanical stability. The complete sample holder has a compact footprint of approximately \SI{2}{\cm} in length, allowing two holders to be installed simultaneously inside the dilution refrigerator. The mating sleeve is mechanically clamped to the cold plate using brass screws, additional copper strands can be wrapped around the ferrules to enhance thermal conductivity.\\

\par\noindent\hspace*{1.5em}\textit{Optical Setup and Cryogenic Environment:} We employ a Sirah Matisse 2DX dye laser to resonantly excite the molecular complexes at wavelengths between \SIrange{579}{581}{\nano\meter}, addressing the coherent $\mathrm{}^7\mathrm{F}_0\rightarrow \mathrm{}^5\mathrm{D}_0$ transition of the \ce{Eu^{3+}} ions. The laser exhibits a FWHM linewidth below \SI{50}{\kilo\hertz} and can be tuned mode-hop-free over a range of \SI{\approx75}{\giga\hertz}. The output of the laser is split into two paths. One beam serves as a reference and is monitored using an optical spectrum analyzer (OSA, High finesse WS6-200) to verify the laser frequency and ensure single-mode operation. The second beam is routed through an acousto-optic modulator (AOM, Gooch \& Housego 3200-121) in a double-pass configuration, enabling precise control of the optical pulse parameters such as amplitude and frequency. The AOM is driven by an arbitrary waveform generator (AWG, Quantum Machines OPX$+$). After modulation, the light is fiber-coupled and directed to the ferrule sample holder inside a dilution refrigerator (Qinu Version LPO). The cryostat is optimized for optical experiments, providing optical access via free-space windows or fiber feedthroughs and enabling operation at millikelvin temperatures. In the measurements presented here, optical access was provided exclusively via the fiber-based ferrule setup. The optical signal transmitted through the microcrystalline samples and the collection fiber of the sample holder is detected using an avalanche photodetector (APD, Thorlabs APD 132A2/M). Selective detection of optical signals is achieved using additional filters. A longpass filter (BLP01-594R-25) is employed to suppress the excitation light and collect the fluorescence, while a bandpass filter (FBH580-10, \SI{\pm10}{\nano\meter} bandwidth) is used for absorption measurements near the resonantly excited $\mathrm{}^7\mathrm{F}_0\rightarrow \mathrm{}^5\mathrm{D}_0$ transition.\\ 

\par\noindent\hspace*{1.5em}\textit{Ensemble Spectroscopy Measurements:} SHB measurements were performed using a two-pulse sequence consisting of a \SI{500}{\milli\second} burn pulse followed by a time-delayed \SI{20}{\milli\second} probe pulse, which was scanned over a frequency range of \SI{60}{\mega\hertz}. All measurements were carried out at low optical powers and without averaging in order to avoid power broadening. The SHB spectra were background-corrected for the frequency-dependent diffraction efficiency of the AOM. In Figure~\ref{fig:F2}(A), the individual spectral holes are vertically offset by a random amount for clarity. To measure the hyperfine state lifetime $T_{1,\text{s}}$, an additional erasing sequence was applied after the probe pulse. This sequence consists of repeated frequency sweeps over a wide spectral range of \SI{160}{\mega\hertz} in order to repopulate the hyperfine levels and restore the initial population distribution prior to the next measurement. For these measurements, a higher optical power of about \SI{1}{\milli\watt} was used to maintain sufficient signal contrast at longer delay times $\tau$ between the burn and probe pulses.\\
FID measurements were conducted using a two-pulse sequence consisting of an initial $\pi/2$ pulse with a duration between \SIrange{2}{4}{\micro\second}, which prepares a coherent superposition of optical ground and excited states, followed by a heterodyne readout pulse used to detect the resulting coherent oscillations of the excited ion ensemble. The readout pulse duration was \SI{2}{\micro\second} and was frequency detuned by \SI{30}{\mega\hertz} relative to the $\pi/2$ pulse for both molecular complexes. The optical powers used for the excitation and readout were \SI{70}{\milli\watt} for Eu-mono and \SI{100}{\milli\watt} for Eu-di. In Figure~\ref{fig:F3}(A), the FID traces are vertically offset by a random amount for clarity.\\ 
Photon echo measurements were performed using a two-pulse echo sequence with heterodyne detection. Each plotted echo amplitude corresponds to a single photon echo measurement at the respective delay time $\tau$ and was obtained by Fast-Fourier transforming the acquired beating signal. Each data point represents an average over 100 pulse sequences. The pulse durations were individually optimized for each molecular complex and ranged between \SIrange{1}{4}{\micro\second}, with optical powers up to \SI{50}{\milli\watt}. The heterodyne detection pulse was detuned by \SI{30}{\mega\hertz} relative to the $\pi/2$ and $\pi$ pulses.\\
Control-target interaction measurements were conducted by adding a frequency-detuned control pulse to the conventional two-pulse photon echo sequence. The delay times $\tau$ between the $\pi/2$, $\pi$, and heterodyne pulses were kept fixed, while the temporal position of the control pulse was scanned between the $\pi$ pulse and the heterodyne readout, defining the evolution time $\tau_{\text{e}}$. In this configuration, the echo amplitude of the target ions is progressively reduced by excitation of the control ions and is plotted as a function of the evolution time, resulting in an exponential decay. The control pulse was detuned by \SI{20}{\mega\hertz} relative to the $\pi/2$ and $\pi$ pulses and applied at optical powers of about \SI{50}{\milli\watt} and \SI{25}{\milli\watt} for Eu-mono and Eu-di, respectively. For Eu-mono, the delay time was fixed to $\tau=\SI{6.5}{\micro\second}$, with pulse durations of \SI{2}{\micro\second} for the $\pi/2$, $\pi$, and heterodyne readout pulses, and \SI{1,5}{\micro\second} for the control pulse. For Eu-di, the delay time was $\tau=\SI{3.4}{\micro\second}$, with pulse durations of \SI{1.7}{\micro\second} for the $\pi/2$, $\pi$, and heterodyne readout pulses, and \SI{1,3}{\micro\second} for the control pulse.\\

\par\noindent\hspace*{1.5em}\textit{Cavity Performance and Integration of a Eu-di-Doped Thin-Film:} The cavity stage is mounted into a liquid helium flow cryostat with a base temperature of \SI{3,5}{\kelvin}. The cavity consists of a single-mode optical fiber with a concave profile of radius of curvature \SI{15}{\micro\meter}, machined onto the end facet by ablation with a \ce{CO2} laser. The machined profile is then coated with a distributed Bragg reflector (DBR) coating with a transmission of \SI{25}{\ppm} at \SI{580}{\nano\meter}. Together with a second DBR planar mirror with a transmission of \SI{200}{\ppm} at \SI{580}{\nano\meter}, a tunable microcavity with a few \si{\micro\meter} air gap is formed.\\ 
A doped thin-film was prepared on the planar mirror. For this purpose, \SI{2.0}{\milli\gram} of Eu-di was dissolved in \SI{3.0}{\milli\liter} of a PMMA solution (E-Beam Resist PMMA 950K AR-P 672.045). The solution was then drop-cast onto the planar mirror, which was subsequently heated in air at \SI{150}{\celsius} for \SI{1}{\hour}, resulting in a thin-film of thickness \SI{\sim2}{\micro\meter} and an average surface roughness of \SI{\sim0.6}{\nano\meter} root-mean-square (RMS) over a $\SI{40}{}\times\SI{40}{\square\micro\meter}$ area (significantly larger than the cavity mode area of \SI{\sim5}{\square\micro\meter}). At this concentration, we expect \SI{\sim e6}{} ions within the cavity mode volume. Due to the relative inhomogeneous and homogeneous broadening determined for this sample (see below), we expect only a subset of $\SI{\sim e3}{}$ ions to contribute to the measured lifetimes when on resonance with the peak of the inhomogeneous line (\SI{580.04}{\nano\meter}).\\
The resulting finesse of the cavity in the presence of the thin-film was measured to be $21{,}500 \pm 100$, higher than the expected finesse of the empty cavity ($17{,}500$) due to the refractive index of the thin-film ($n=1.49$ at \SI{580}{\nano\meter}) more closely matching that of \ce{SiO2} used in the DBR coating. For the measurements presented here, the cavity length was actively stabilized with an RMS fluctuation of \SI{\sim9}{\pico\meter}.\\
The inhomogeneous linewidth of Eu-di in the PMMA thin-film was measured to be \SI{1.2}{\nano\meter}, centered at \SI{580.04}{\nano\meter} with a homogeneous linewidth of \SI{\sim200}{\mega\hertz} at \SI{4}{\kelvin}.

\newpage

\medskip
\textbf{Supporting Information} \par
Supporting Information is available in a separate PDF file, which contains:\\
Synthetic Procedures\\
Additional Characterization\\
Quantum Efficiency Calculation\\
Dipole--Dipole Interaction Estimates for Eu-mono and Eu-di\\
Figures S1 to S3\\
Tables S1 to S2\\

\newpage
\medskip
\textbf{Acknowledgements}\par
E. V., V. U. Ch., B. B., N. L. J., S. K. K., M. R. and D. H. acknowledge funding from the Deutsche Forschungsgemeinschaft (DFG) through the Collaborative Research Centre “4f for Future” (CRC 1573 project number 471424360, project C2), the Max-Planck-School of Photonics, the Karlsruhe School of Optics and Photonics (KSOP). M. R. acknowledges funding from the government programme managed by the French National Research Agency under grants ANR-20-CE09-0022-01 (UltraNanoSpec), ANR-23-CE47-0011 (MoleQuBe) and France 2030-ANR-MOlQif (PEPR Technologies Quantiques).\\
The authors acknowledge the use of ChatGPT (accessed march 2026) and Microsoft Copilot (accessed March 2026) for stylistic language editing and polishing of the manuscript. All content was reviewed and approved by the authors, who take full responsibility for the accuracy and integrity of the work.\linebreak 

\textbf{Author Contributions}\par
E. V. and V. U. Ch. set up the experiment and conducted the measurements. B. B. and S. K. K. proposed and synthesized both molecular complexes. N. L. J. performed the cavity experiments on the dinuclear complex. M. R. and D. H. advised on all efforts. All authors contributed to the data analysis and the manuscript preparation.\linebreak

\textbf{Conflict of Interest}\par
The authors declare no conflict of interest.\linebreak

\textbf{Data Availability Statement}\par
The data that support the findings of this study are available from the corresponding author upon reasonable request.

\newpage

\medskip







\end{document}



\pagestyle{fancy}
\fancyhf{}
\fancyfoot[C]{\thepage}

\title{Supporting Information\\[0.5em]
Controlled ion--ion interactions and cavity-enhanced emission of a coherent dinuclear Eu\textsuperscript{3+} complex}

\maketitle

\author{Evgenij Vasilenko\equalcontrib, Vishnu Unni Chorakkunnath\equalcontrib, Barbora Brachnakova\equalcontrib, Nicholas Lester Jobbitt, Senthil Kumar Kuppusamy, David Hunger*, and Mario Ruben*}

\dedication{$^{\dagger}$\ These authors contributed equally to this work.}


\begin{affiliations}

Evgenij Vasilenko, Barbora Brachnakova, Senthil Kumar Kuppusamy, David Hunger, Mario Ruben\\
Karlsruhe Institute of Technology, Institute for Quantum Materials and Technologies (IQMT), Eggenstein-Leopoldshafen, 76344, Germany\\
E-mail: \texttt{mario.ruben@kit.edu; david.hunger@kit.edu}\linebreak

Evgenij Vasilenko, Vishnu Unni Chorakkunnath, Nicholas Lester Jobbitt, David Hunger\\
Karlsruhe Institute of Technology, Physics Institute (PHI), Wolfgang-Gaede-Straße 1, Karlsruhe, 76131, Germany\linebreak

Mario Ruben\\
Karlsruhe Institute of Technology, Institute of Nanotechnology (INT), Eggenstein-Leopoldshafen, 76344, Germany\linebreak

Mario Ruben\\
Centre Europ\'{e}en de Sciences Quantiques (CESQ), Institut de Science et d’Ing\'{e}nierie Supramol\'{e}culaires (ISIS), Strasbourg, 67083, France\\

\end{affiliations}

\newpage


\tableofcontents


\renewcommand{\thefigure}{S\arabic{figure}}
\renewcommand{\thetable}{S\arabic{table}}
\renewcommand{\theequation}{S\arabic{equation}}
\renewcommand{\thepage}{S\arabic{page}}
\setcounter{figure}{0}
\setcounter{table}{0}
\setcounter{equation}{0}
\setcounter{page}{1}

\newpage


\section{Synthetic Procedures}
\label{sec:S1}

\subsection{Synthesis of the Precursor}

First, the precursor \ce{[Eu(btfa)3(H2O)2]} was prepared according to the following procedure. A mixture of 4,4,4-trifluoro-1-phenyl-1,3-butanedione (\SI{3}{\milli\mol}, \SI{650}{\milli\gram}, 3 eq) dissolved in \SI{10}{\milli\liter} of ethanol (\SI{96}{\percent}) and \SI{1}{\milli\liter} of an aqueous solution of \ce{NaOH} (\SI{3}{\milli\mol}, \SI{120}{\milli\gram}, 3 eq) was stirred at room temperature for \SI{2}{\hour}. Subsequently, \SI{2}{\milli\liter} of an ethanolic solution of \ce{EuCl3.6H2O} (\SI{1}{\milli\mol}, \SI{366}{\milli\gram}, 1 eq) was added, and the mixture was stirred for an additional \si{\hour}. After adding \SI{20}{\milli\liter} of water, the resulting white precipitate was collected by filtration, washed with a small volume of cold ethanol, and dried under vacuum for \SI{48}{\hour}. Elemental analysis for \ce{C30H22O8F9Eu}: C (calcd/found) 43.10/43.29; H (calcd/found) 3.00/2.68.

\subsection{Synthesis of Eu-mono}

A mixture of 4,4,4-trifluoro-1-phenyl-1,3-butanedione (\SI{3}{\milli\mol}, \SI{650}{\milli\gram}, 3 eq) dissolved in \SI{5}{\milli\liter} of ethanol (\SI{96}{\percent}) and \SI{2}{\milli\liter} of an aqueous solution of \ce{NaOH} (\SI{3}{\milli\mol}, \SI{120}{\milli\gram}, 3 eq) was stirred at room temperature for \SI{2}{\hour}. Subsequently, \SI{2}{\milli\liter} of an ethanolic solution of \ce{EuCl3.6H2O} (\SI{1}{\milli\mol}, \SI{366}{\milli\gram}, 1 eq) and \SI{1}{\milli\liter} of an ethanolic solution of 2,2´-bipyridine (\SI{1}{\milli\mol}, \SI{156}{\milli\gram}, 1 eq) were added, and the mixture was stirred for an additional \SI{1}{\hour}. The resulting white precipitate was filtered, washed with a small volume of water and ethanol, and dried under vacuum for \SI{48}{\hour}. Recrystallization of \SI{50}{\milli\gram} of the microcrystalline powder from a \SI{5}{\milli\liter} mixture of water and ethanol (1:2) yielded large, colorless, rhomboid-shaped crystals in \SI{86}{\percent} yield. Elemental analysis for \ce{C40H26O6N2F9Eu}: C (calcd/found) 50.20/50.42; H (calcd/found) 2.75/2.79; N (calcd/found) 2.94/2.90.

\subsection{Synthesis of Eu-di}
\sloppy
The dinuclear complex \ce{[Eu2(btfa)6(bpim)]} was prepared by dissolving the corresponding precursor \ce{[Eu(btfa)3(H2O)2]} (\SI{0,12}{\milli\mol}, \SI{100}{\milli\gram}, 2 eq) and 2,2'-bipyrimidine (\SI{0,06}{\milli\mol}, \SI{9,5}{\milli\gram}, 1 eq) in \SI{3}{\milli\liter} of ethanol (\SI{96}{\percent}). Slow solvent evaporation led to the formation of needle-like crystals within 5 days in \SI{68}{\percent} yield. Elemental analysis for \ce{C68H42O12N4F18Eu2} : C (calcd/found) 46.59/46.59; H (calcd/found) 2.41/2.43; N (calcd/found) 3.20/3.18.

\begin{figure}[tbp] 
\centerline{\includegraphics[width=\columnwidth]{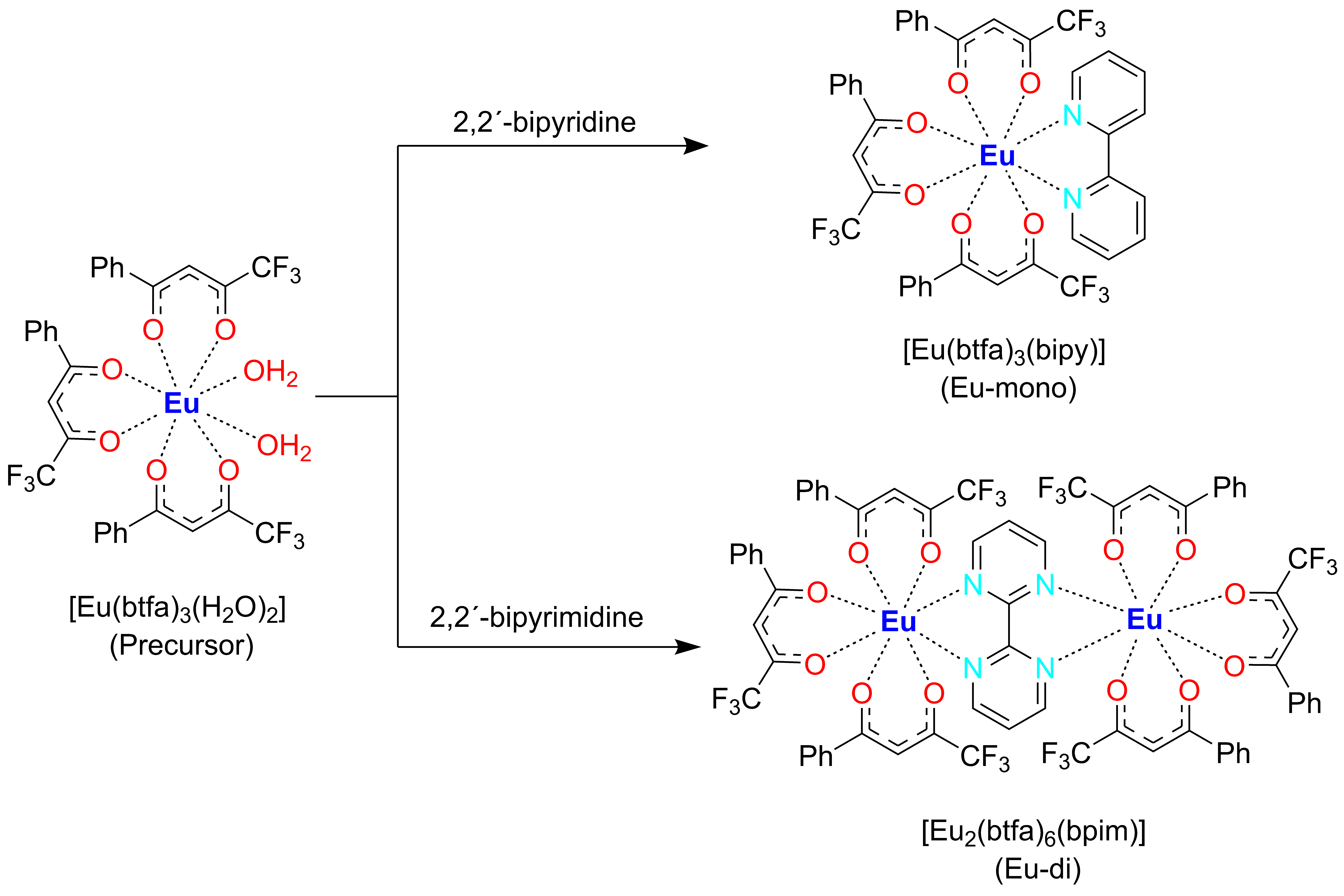}}
\caption{\textbf{Schematic synthesis of mono- and dinuclear complexes.} The precursor (left) yields the mononuclear complex (Eu-mono, upper right) and the dinuclear complex (Eu-di, lower right) upon coordination with the corresponding N-donor ligands indicated.}
\label{fig:SI_1}
\end{figure}

The synthetic procedure and respective molecular components are summarized schematically in Figure~\ref{fig:SI_1}, illustrating the conversion of the precursor into the mononuclear (Eu-mono) and dinuclear (Eu-di) complexes via coordination with 2,2'-bipyridine and 2,2'-bipyrimidine, respectively.

\section{Additional Characterization}
\label{sec:S2}

\subsection{Structural Details}

\begin{figure}[tbp] 
\centerline{\includegraphics[width=\columnwidth]{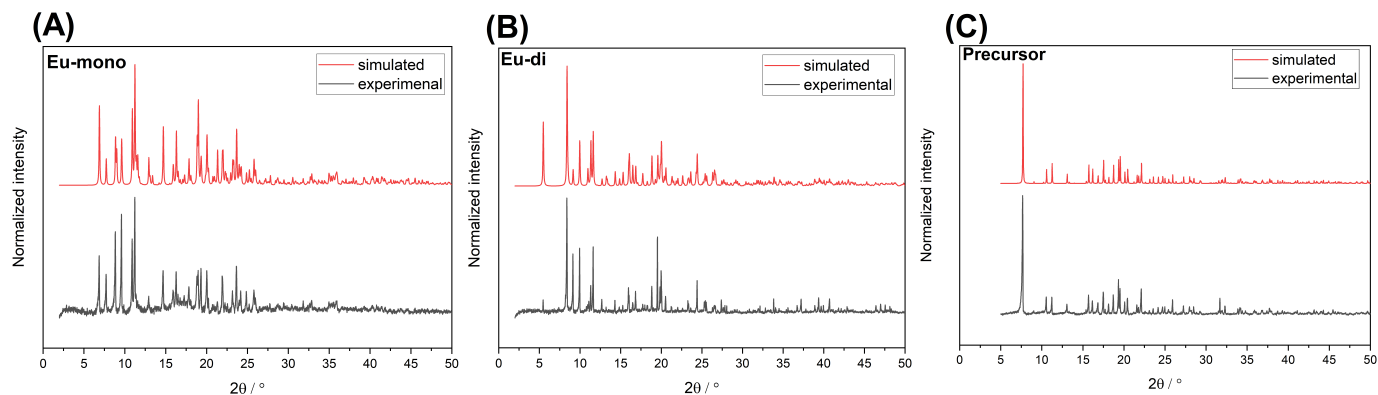}}
\caption{\textbf{Simulated (red) and experimental (black) powder X-ray diffraction patterns of polycrystalline samples recorded at \SI{293}{\kelvin}. (A):} Eu-mono, \textbf{(B):} Eu-di, and \textbf{(C):} the precursor.}
\label{fig:SI_2}
\end{figure}

The coordination sphere of the \ce{Eu^{3+}} center of the asymmetric unit in Eu-mono and Eu-di consists of six oxygen atoms from three coordinating 4,4,4-trifluoro-1-phenyl-1,3-butanedione ligands and two nitrogen atoms from the bidentate 2,2’-bipyridine and 2,2’-bipyrimidine ligands, respectively. Because both structures have been previously reported \cite{Batista, Ilmi, van1996synthesis}, single-crystal and powder X-ray diffraction analyses were used to determine the unit cell parameters of the crystals and to confirm the phase purity of the microcrystalline materials used for PL, SHB and photon echo studies. The experimental patterns of the precursor, Eu-mono, and Eu-di are in good agreement with the simulated patterns calculated from the CIF files, as shown in Figure~\ref{fig:SI_2}.\\

Eu-mono possesses a monoclinic structure with the $P2_1/n$ space group, while \ce{Eu-di} crystallizes in a triclinic system with the $P\bar{1}$ space group. The Eu--O bond distances are in the range between \SIrange{2.321}{2.399}{\angstrom} for both complexes, and the values are comparable to other mono- and dinuclear complexes based on $\beta$-diketonate \cite{DESILVA20073543}. The distances Eu--$\mathrm{N}_1$ and Eu--$\mathrm{N}_2$ are approximately \SI{0.05 \pm 0.01}{\angstrom} and \SI{0.07 \pm 0.02}{\angstrom} longer in Eu-di due to reduced electron density and the lower basicity of the nitrogen atoms in 2,2’-bipyrimidine resulting from the electron-withdrawing effect of the additional nitrogen atoms in the aromatic rings \cite{zucchi2011utility}. Additionally, the coordination of N-donor ligands facilitates structural assembly via short contacts in Eu-mono and coordination bonds in Eu-di. In the mononuclear complex, the shortest Eu--Eu distance between two asymmetric units is \SI{8,459}{\angstrom}, which arises from weak $\pi$--$\pi$ stacking between neighboring bipyridine ligands (centroid--centroid distance of \SI{3,89}{\angstrom}). A shorter Eu--Eu distance (\SI{6,914}{\angstrom}) is found within the molecular complex Eu-di, reflecting a more rigid structural arrangement that can enhance the electronic coupling between the two \ce{Eu^{3+}} centers. This in turn may stabilize the excited emissive state, contributing to a more intense and well-defined $0\rightarrow0$ transition in the PL spectrum.\\

To gain further insight into the coordination symmetry around \ce{Eu^{3+}} centers, we conducted continuous shape measure (CShM) analyses \cite{pinsky1998continuous}, the results of which are summarized in Table~\ref{tab:complex_data}. Calculated CShM values greater than zero confirm a deviation from ideal symmetry. The lowest calculated values are 0.647 for Eu-mono and 0.951 for both centers in Eu-di, indicating the square antiprismatic coordination geometry with $D_{4d}$ symmetry. In contrast, the precursor has a higher value of 2.958, indicating that replacement of aqua ligands by N-donor ligands enhances geometric symmetry. On the other hand, TDD-8 parameter shows the opposite trend: the precursor is nearly ideal (CShM$=0.193$), while distortion increases in the complexes (CShM$=2.472$ for Eu-mono and $1.535$ for Eu-di), which likely reduces the lifetime and causes spectral broadening in the precursor. These findings emphasize that improving symmetry, especially in highly favorable geometries like $D_{4d}$, is critical for optimizing PL properties.

\begin{table}[b]
\captionsetup{justification=raggedright,singlelinecheck=false}
\centering
\caption{Continuous shape measure values quantifying the deviation of the coordination polyhedra from ideal geometries for Eu-mono, Eu-di, and the precursor.}
\label{tab:complex_data}
\begin{tabular}{|c|c|c|c|c|c|}
\hline
 \textbf{Code} & \textbf{Symmetry} & \textbf{Geometry} & \textbf{Eu-mono} & \textbf{Eu-di} & \textbf{Precursor} \\
\hline
    OP-8     & $D_{8h}$  & Octagon                             & 30.141 & 30.910 & 33.527 \\
    HPY-8    & $C_{7v}$  & Heptagonal pyramid                  & 22.285 & 22.817 & 25.063 \\
    HBPY-8   & $D_{6h}$  & Hexagonal bipyramid                 & 15.404 & 16.060 & 14.581 \\
    CU-8     & $O_{h}$   & Cube                                & 9.285  & 9.551  & 7.010  \\
    SAPR-8   & $D_{4d}$  & Square antiprism                    & 0.647  & 0.951  & 2.958  \\
    TDD-8    & $D_{2d}$  & Triangular dodecahedron             & 2.472  & 1.535  & 0.193  \\
    JGBF-8   & $D_{2d}$  & Johnson gyrobifastigium J26         & 15.203 & 15.535 & 15.559 \\
    JETBPY-8 & $D_{3h}$  & Johnson elongated triangular bipyramid        & 27.600 & 27.600 & 29.373 \\
    JBTPR-8  & $C_{2v}$  & Biaugmented trigonal prism J        & 2.797  & 2.800  & 3.410  \\
    BTPR-8   & $C_{2v}$  & Biaugmented trigonal prism          & 2.118  & 2.250  & 3.031  \\
    JSD-8    & $D_{2d}$  & Snub diphenoid J84                  & 4.845  & 4.394  & 2.980  \\
    TT-8     & $T_{d}$   & Triakis tetrahedron                 & 10.027 & 10.311 & 7.620  \\
    ETBPY-8  & $D_{3d}$  & Elongated trigonal bipyramid        & 23.308 & 23.447 & 25.382 \\
   \hline 
    \bottomrule
  \end{tabular}
\end{table}

\subsection{Photophysical Properties}

\begin{figure}[tbp] 
\centerline{\includegraphics[width=\columnwidth]{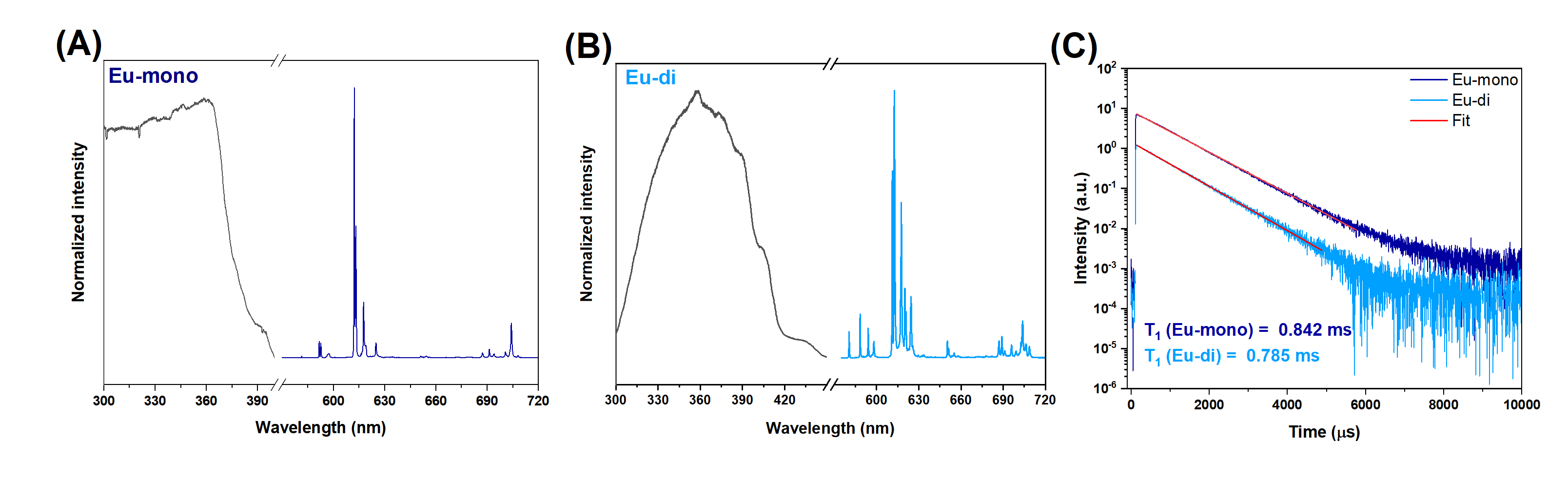}}
\caption{\textbf{Photoluminescence spectroscopy of Eu-mono and Eu-di at \SI{3}{\kelvin}. (A):} Excitation spectrum (black) recorded at the emission maximum at \SI{612}{\nano\meter} and scanned over \SIrange{300}{400}{\nano\meter}, together with the emission spectrum recorded upon excitation at \SI{365}{\nano\meter} in the wavelength range \SIrange{550}{750}{\nano\meter} for Eu-mono. \textbf{(B):}Same as (A) for Eu-di, with excitation wavelengths scanned over \SIrange{300}{450}{\nano\meter}. \textbf{(C):} Time-resolved PL decay curves of Eu-mono and Eu-di monitored at \SI{612}{\nano\meter}.}
\label{fig:SI_3}
\end{figure}

High-resolution PL experiments were performed at \SI{3}{\kelvin} using a crystalline sample (see Figure~\ref{fig:SI_3}). Emission spectra were collected upon excitation at \SI{365}{\nano\meter} (see Figure~\ref{fig:SI_3}(A) and (B)), with the slit fully open (\SI{7}{\nano\meter}) to ensure sufficient excitation intensity and the emission slit set to \SI{0,2}{\nano\meter} to achieve high spectral resolution. The emission peaks observed in the range \SIrange{550}{750}{\nano\meter} correspond to the characteristic \ce{Eu^{3+}} transitions from the $\mathrm{}^5\mathrm{D}_0$ excited state to the $\mathrm{}^7\mathrm{F}_J$ ($J$=0-4) ground state manifold. The narrow $\mathrm{}^5\mathrm{D}_0\rightarrow\mathrm{}^7\mathrm{F}_0$ transition located around \SI{580}{\nano\meter} is observed in both molecular systems and is of particular interest in this study. Theoretically, in an ideal $D_{4d}$ site symmetry this transition is formally forbidden. However, a single emission line is experimentally detected in both complexes, indicating the presence of at least one local \ce{Eu^{3+}} site lacking an inversion center, consistent with the crystallographic analysis. In the dinuclear system, the inversion center is located on the bridging ligand, while each \ce{Eu^{3+}} site individually lacks inversion symmetry, allowing the $0\rightarrow0$ transition.\\

The $\mathrm{}^5\mathrm{D}_0\rightarrow\mathrm{}^7\mathrm{F}_1$ transition is of magnetic dipole character with moderate intensity around \SI{590}{\nano\meter} and for lower symmetries, a splitting into at least three sublevels is expected which is consistent with our observations. In contrast, the $\mathrm{}^5\mathrm{D}_0\rightarrow\mathrm{}^7\mathrm{F}_2$ transition dominates the emission spectra, contributing more than \SI{70}{\percent} of the total integrated intensity in both complexes. This transition is highly sensitive to the local environment of \ce{Eu^{3+}} sites and shows a pronounced splitting, with five or more peaks observed in the region between \SIrange{610}{630}{\nano\meter}. Similar spectral characteristics have been reported for \ce{Eu^{3+}} coordination compounds \cite{Binnemans}. Furthermore, the relative intensity ratio of $\mathrm{}^5\mathrm{D}_0\rightarrow\mathrm{}^7\mathrm{F}_2$ to $\mathrm{}^5\mathrm{D}_0\rightarrow\mathrm{}^7\mathrm{F}_1$ is commonly used as an indicator of the asymmetry of the \ce{Eu^{3+}} coordination environment. In general, \ce{Eu^{3+}} complexes with a centrosymmetric coordination sphere exhibit an intensity ratio of less than \SI{0,7}{}. Conversely, a ratio exceeding \SI{8}{} suggests a highly asymmetric coordination environment around the \ce{Eu^{3+}} ion. The calculated intensity ratios are \SI{12,5}{} and \SI{10,9}{} for the Eu-mono and Eu-di, respectively. The introduction of N-donor ligands enhances not only the PL intensity but also indicates that the coordination polyhedron is more distorted from the ideal $D_{4d}$ symmetry towards to lower symmetries, as calculated by the CShM method. The emission decay curves of both \ce{Eu^{3+}} complexes were monitored at \SI{612}{\nano\meter} upon UV excitation. The observed excited state lifetime $T_{1,\text{o}}$ was extracted by monoexponential fitting of the time-dependent PL decay curves (see Figure~\ref{fig:SI_3}(C)), and the obtained values are comparable to those previously reported  \cite{ebert2025photoluminescence, Ilmi}.\\

Table~\ref{tab:BR} shows the branching ratios $\beta_{J}$ for the $\mathrm{}^5\mathrm{D}_0\rightarrow\mathrm{}^7\mathrm{F}_J$ transitions ($J$=0-4), calculated from the experimental emission spectra as the ratio of the integrated intensity of each transition within the selected spectral range to the total integrated intensity over all transitions. The calculations were performed using the LUMPAC software \cite{dutra2014lumpac}.

\begin{table}[tbp]
\caption{Calculated branching ratios $\beta_J$ for $\mathrm{}^5\mathrm{D}_0\rightarrow\mathrm{}^7\mathrm{F}_J$ transitions within the selected spectral ranges for Eu-mono and Eu-di.}
\label{tab:BR}
\centering
\begin{tabular}{|c|c|c|c|c|}
\hline
 & \multicolumn{2}{c|}{\textbf{Eu-mono}} & \multicolumn{2}{c|}{\textbf{Eu-di}} \\ \hline
Transition & Selected range \si{\nano\meter} &  $\beta_J$ (\si{\percent}) 
& Selected range (\si{\nano\meter}) & $\beta_J$ (\si{\percent})  \\ \hline

$\mathrm{}^5\mathrm{D}_0\rightarrow\mathrm{}^7\mathrm{F}_0$
& 581--582 & 0.14 
& 579.5--581.5 & 1.30 \\ \hline

$\mathrm{}^5\mathrm{D}_0\rightarrow\mathrm{}^7\mathrm{F}_1$
& 591--599 & 5.91 
& 587--601 & 6.48 \\ \hline

$\mathrm{}^5\mathrm{D}_0\rightarrow\mathrm{}^7\mathrm{F}_2$
& 610--626 & 74.22 
& 609--628 & 70.47 \\ \hline

$\mathrm{}^5\mathrm{D}_0\rightarrow\mathrm{}^7\mathrm{F}_3$
& 647--660 & 1.66 
& 648--662 & 3.07 \\ \hline

$\mathrm{}^5\mathrm{D}_0\rightarrow\mathrm{}^7\mathrm{F}_4$
& 685--710 & 18.08 
& 685--711 & 18.68 \\ \hline

\end{tabular}
\end{table}


\section{Quantum Efficiency Calculation}
\label{sec:S3}

The quantum efficiency of the Eu-di thin-film is determined by comparing the observed free-space lifetime with the radiative lifetime reported in Ref.~\cite{Ilmi}, and is defined as their ratio:
\begin{equation}\label{eq:quantum_eff}
    \mathrm{QE}= \frac{T_{\text{1,o}}}{T_{\text{1,rad}}}=\frac{\SI{509}{\micro\second}}{\SI{1072}{\micro\second}}.
\end{equation}
From this,we obtain a $\mathrm{QE}$ of $\SI{\sim 0,480}{}$.


\section{Dipole--Dipole Interaction Estimates for Eu-mono and Eu-di}
\label{sec:S4}

Controlled interaction experiments rely on the change in permanent electric dipole moment of \ce{Eu^{3+}} ions upon optical excitation. When a control ion $i$ is excited, the resulting change in its electric dipole moment produces a frequency shift of a nearby target ion $j$ through dipole-dipole interactions. The interaction-induced frequency shift at a distance $r_{ij}$ can be expressed as \cite{Ahlefeldt_2013}:  
\begin{equation}\label{eq:freq_shift}
    \Delta f_{ij}=\frac{\Delta\mu_{\text{$i$,eff}}\Delta\mu_{\text{$j$,eff}}}{4\pi\epsilon\epsilon_0\, h\, r_{ij}^3}\underbrace{[\mathbf{\hat{\mu}}_i\cdot\mathbf{\hat{\mu}}_j-3(\mathbf{\hat{\mu}}_i\cdot\mathbf{\hat{r}}_{ij})(\mathbf{\hat{\mu}}_j\cdot\mathbf{\hat{r}}_{ij})]}_{\Theta} ,
\end{equation}
where $\Delta\mu_{\text{eff}}$ is the effective difference between ground and excited state dipole moments, $\epsilon_0$ is the vacuum permittivity, and $\epsilon$ the dielectric constant of the host material. The vector $\mathbf{\hat{r}}_{ij}$ denotes the unit vector connecting ions $i$ and $j$, while $\mathbf{\hat{\mu}}_i$ and $\mathbf{\hat{\mu}}_j$ describe the relative orientations of the dipole moment changes upon optical excitation and dipole-dipole interaction. The latter term can be summarized by the geometric factor $\Theta$, which ranges from $-2$ to 1.\\

To estimate the interaction strength in the Eu-mono and Eu-di complexes, we consider the shortest Eu--Eu separations obtained from the crystal structures. The Eu-mono unit cell contains four molecules with lattice parameters $a=\SI{11,122}{\angstrom}$, $b=\SI{22,860}{\angstrom}$, and $c=\SI{15,870}{\angstrom}$. Within the unit cell, Eu--Eu separations range from \SI{8,459}{\angstrom} up to approximately \SI{21,038}{\angstrom}. For Eu-di, the unit cell contains one molecule with lattice parameters $a=\SI{9,800}{\angstrom}$, $b=\SI{11,080}{\angstrom}$, and $c=\SI{17,150}{\angstrom}$. The shortest Eu--Eu separation within the unit cell is the intramolecular distance of \SI{6,914}{\angstrom}, while distances to \ce{Eu^{3+}} ions in neighboring molecules range from \SI{9,800}{\angstrom} up to about \SI{17,150}{\angstrom}.\\ 

Assuming a geometric factor of $\Theta=1$, corresponding to parallel dipole moments oriented perpendicular to the $\mathbf{\hat{r}}_{ij}$ vector, and an effective dipole moment on the order of \SI{7e-32}{\coulomb\meter} (\SI{1,04e-32}{\coulomb\meter} \cite{Ahlefeldt_2013},\SI{7,326e-32}{\coulomb\meter} \cite{Graf_1998}, and \SI{3,31e-31}{\coulomb\meter} \cite{serrano_ultra_2022}), we estimate the interaction-induced frequency shift using Equation~\ref{eq:freq_shift}. Since the dielectric constant of both molecular complexes is not known, we approximate it using the refractive index $n\approx 1.5$ \cite{serrano_ultra_2022}, yielding $\epsilon\approx n^2=2.25$. Together, using the shortest Eu--Eu separations within the crystal structures $r_{\text{mono}}=\SI{8,459}{\angstrom}$ for Eu-mono and $r_{\text{di}}=\SI{6,914}{\angstrom}$ for Eu-di, the corresponding frequency shifts are estimated to be on the order of:
\begin{equation}
    \Delta f_{\text{mono}} \approx \SI{50}{\mega\hertz}, \qquad
    \Delta f_{\text{di}} \approx \SI{90}{\mega\hertz}.
\end{equation}
These values represent estimates of maximal values corresponding to the closest ion pairs and favorable dipole orientation, while larger Eu--Eu separations within the crystal lead to substantially smaller interaction-induced frequency shifts due to the $1/r^3$ dependence of the dipole-dipole interaction.\\

For a randomly distributed ensemble of control and target ions, excitation of control ions leads to an effective broadening of the optical transition of the target ions. This excitation-induced broadening can be approximated by \cite{Altner_1996}:   
\begin{equation}\label{eq:prob_exc}
    \Gamma_{\text{c}}=\frac{2}{3}\pi^2
    \frac{(\Delta\mu_{\text{eff}})^2}{4\pi\epsilon\epsilon_0\, h}
    \rho_0 p,
\end{equation}
where $\rho_0$ denotes the density of control ions and $p$ the fraction of the ions excited by the control pulse.\\

From the controlled interaction experiments, we obtained an interaction-induced broadening of $\Gamma_{\text{c}}=\SI{6}{\kilo\hertz}$ for Eu-mono and $\Gamma_{\text{c}}=\SI{19}{\kilo\hertz}$ for Eu-di. The \ce{Eu^{3+}} ion density $\rho_{\text{tot}}$ can be estimated from the crystal structures using the unit cell volumes determined from the lattice parameters given above. For Eu-mono, the unit cell volume is \SI{4,04e-21}{\cubic\centi\meter} containing four \ce{Eu^{3+}} ions, resulting in a total ion density $\rho_{\text{tot,mono}}\approx \SI{9.9e20}{\per\cubic\centi\meter}$. For Eu-di, the unit cell volume is \SI{1.87e-21}{\cubic\centi\meter} containing two \ce{Eu^{3+}} ions, corresponding to a total ion density $\rho_{\text{tot,di}}\approx \SI{1.1e21}{\per\cubic\centi\meter}$. These values refer to the total \ce{Eu^{3+}} densities in the crystals and therefore represent an upper bound for the density of control ions participating in the control-target experiment.\\

Assuming the same effective dipole moment of $\Delta\mu_{\text{eff}}\approx \SI{7e-32}{\coulomb\meter}$, the fraction of control ions excited by the control pulse can then be estimated from Equation~\ref{eq:prob_exc}. Using the experimentally determined broadenings $\Gamma_{\text{c}}$ and the total \ce{Eu^{3+}} ion densities calculated above, we obtain excitation fractions of $p_{\text{mono}}\approx\SI{6,8e-7}{}$ and $p_{\text{di}}\approx\SI{2,0e-6}{}$ for Eu-mono and Eu-di, respectively. This corresponds to densities of excited control ions of:
\begin{equation}
    (\rho_0p)_{\text{mono}} \approx \SI{6,7e14}{\per\cubic\centi\meter}, \qquad
    (\rho_0p)_{\text{di}} \approx \SI{2,2e15}{\per\cubic\centi\meter}.
\end{equation}
These densities indicate that only a very small fraction of ions within the inhomogeneously broadened ensemble is excited during the control pulse, consistent with excitation fractions reported in two-pulse photon echo experiments \cite{Konz_2003}.\\ 

Using the measured coherence times $T_{\text{2,o(mono)}}=\SI{9}{\micro\second}$ and $T_{\text{2,o(di)}}=\SI{4}{\micro\second}$, the corresponding homogeneous linewidths $\Gamma_{\text{h}}=1/\pi T_{\text{2,o}}$ are approximately \SI{35}{\kilo\hertz} and \SI{80}{\kilo\hertz}, respectively. The distance at which the dipole-dipole interaction shift becomes comparable to the homogeneous linewidth yields interaction radii of approximately \SI{9}{\nano\meter} for Eu-mono and \SI{7}{\nano\meter} for Eu-di. Within this interaction volume, each excited control ion therefore influences on the order of \SI{e3}{} surrounding target ions.

\newpage

